\documentclass[12pt]{article}
\pdfoutput=1
\usepackage{graphicx,epstopdf,amssymb,amsfonts,amsmath,amsthm,array,
mathrsfs,amscd}

\newcommand{\pbs}[1]{\let\temp=\\#1\let\\=\temp}
\numberwithin{equation}{section}
\def\be{\begin{equation}}\def\ee{\end{equation}}
\def\cvp{\raise 2pt\hbox{,}}

 \def\d{{\rm d}}

\def\la{\lambda}\def\La{\Lambda}

\def\Re{\mathop{\rm Re}}
\def\d{\partial}

\def\d{{\rm d}}

\def\Det{{\rm Det}}

\def\a{\alpha}
\def\b{\beta}
\def\g{\gamma}
\def\G{\Gamma}
\def\dd{\delta}

\def\m{\mu}

\def\k{\kappa}
\def\l{\lambda}

\def\s{\sigma}
\def\f{\phi}
\def\p{\psi}

\def\D{\Delta}
\def\z{\zeta}

\def\vf{\varphi}
\def\L{\Lambda}

\def\wh{\widehat}
\def\wt{\widetilde}

\def\l{\lambda}

\def\ba{\begin{eqnarray}}
\def\ea{\end{eqnarray}}

\theoremstyle{plain}

\theoremstyle{definition}
\theoremstyle{remark}

\def\npb#1#2#3{{\it Nucl.\ Phys.\ }{\bf B #1} (#2) #3}
\def\npps#1#2#3{{\it Nucl.\ Phys.\ Proc.\ Suppl.\ }{\bf #1} (#2) #3}

\def\prd#1#2#3{{\it Phys.\ Rev.\ }{\bf D #1} (#2) #3}

\def\cmp#1#2#3{{\it Comm.\ Math.\ Phys.\ }{\bf #1} (#2) #3}

\def\jmp#1#2#3{{\it J.\ Math.\ Phys.\ }{\bf #1} (#2) #3}

\def\imath#1#2#3{{\it Invent math }{\bf #1} (#2) #3}

\def\jdiffgeo#1#2#3{{\it J.\ Diff.\ Geom.\ }{\bf #1} (#2) #3}

\begin{document}
%
%
{\pagestyle{empty}
\parskip 0in
\

\vfill
\begin{center}



{\LARGE 2D Quantum Gravity at One Loop}

\medskip

{\LARGE  with Liouville and Mabuchi Actions}

\vspace{0.4in}

Adel B{\scshape ilal}$^{*}$, Frank F{\scshape errari}$^{\dagger}$ and Semyon K{\scshape levtsov}$^{\sharp}$
\\
\medskip
{$^{*}$\it Centre National de la Recherche Scientifique\\
Laboratoire de Physique Th\'eorique de l'\'Ecole Normale Sup\'erieure\\
24 rue Lhomond, F-75231 Paris Cedex 05, France}

\smallskip

{$^{\dagger}$\it Service de Physique Th\'eorique et Math\'ematique\\
Universit\'e Libre de Bruxelles and International Solvay Institutes\\
Campus de la Plaine, CP 231, B-1050 Bruxelles, Belgique}

\smallskip

{$^{\sharp}$\it Mathematisches Institut and Institut f\"ur Theoretische Physik,\\ Universit\"at zu K\"oln, Weyertal 86-90, 50931 K\"oln, Germany

}

\smallskip

{\tt\small adel.bilal@lpt.ens.fr, frank.ferrari@ulb.ac.be, klevtsov@math.uni-koeln.de}
\end{center}
\vfill\noindent
We study a new two-dimensional quantum gravity theory, based on a gravitational action containing both the familiar Liouville term and the Mabuchi functional, which has been shown to be related to the coupling of non-conformal matter to gravity. We compute the one-loop string susceptibility from a first-principle, path integral approach in the
K\"ahler parameterization of the metrics and discuss  the particularities that arise in the case of the pure Mabuchi theory. While we mainly use the most convenient spectral cutoff regularization to perform our calculations, we also discuss the interesting subtleties associated with the multiplicative anomaly in the 
familiar $\zeta$-function scheme, which turns out to have a genuine physical effect for our calculations. In particular, we derive and use a general multiplicative anomaly formula for Laplace-type operators on arbitrary compact Riemann surfaces.

\vfill

\medskip
%
\begin{flushleft}
\today
\end{flushleft}
\newpage\pagestyle{plain}
\baselineskip 16pt
\setcounter{footnote}{0}
\setcounter{page}{1}

}


%
\section{Introduction}
\setcounter{equation}{0}

Two dimensional gravity on Riemann surfaces has been studied since long by various methods. Discrete approaches involve triangulations \cite{triang} and matrix models \cite{matrixold}. The latter also easily incorporate the coupling to various types of conformal matter systems \cite{matrixmatter}. The continuum approach has been mainly focussing on conformal gauge and the coupling to conformal matter using the Liouville action \cite{Liouville1}. While this action is relatively well understood \cite{Liouville2}, the appropriate functional integral measure is complicated, very different from a standard Gaussian one. Nevertheless, it was shown in \cite{DDK} that, if one adopts the simplifying assumption of a free-field measure, one can still and quite remarkably satisfy the fundamental constraint of background independence, at the expense of modifying some parameters in the Liouville action. This approach then yields the celebrated formula for the area dependence of the fixed-area quantum gravity partition function in the presence of conformal matter with central charge $c$ \cite{DDK}. If one defines the coupling constant $\kappa^{2}$ by
\be\label{kappadef} \k^2=\frac{26-c}{3}\, \cvp\ee
one finds that the partition function scales as
\be\label{Zscale} Z(A)\sim e^{\m^2 A} A^{\g_{\rm str}(\kappa^{2})-3}\, ,\ee
where $\mu^{2}$ is a renormalization-dependent cosmological constant and $\g_{\rm str}$ is the string susceptibility for a genus $h$ compact Riemann surface, given by 
\be\label{KPZformula}
\g_{\rm str}(\kappa^{2})= 2+2(h-1)\frac{\sqrt{\k^2-1/3}}{\sqrt{\k^2-1/3}-\sqrt{\k^2-25/3}}\, \cdotp\ee
The semi-classical limit of the above formula corresponds to $c\rightarrow -\infty$ or equivalently to $\k^2\to\infty$. More precisely, expanding in inverse powers of $\k^2$, a term in $\k^{2(1-L)}$ corresponds to the $L$-loop contribution. Thus, up to one loop,
\be\label{gstrexp}
 \g_{\rm str}=\frac{h-1}{2}\k^2+\frac{19-7h}{6} + O(\kappa^{-2}) \, .
 \ee
Notwithstanding its non-rigorous derivation, the formula \eqref{KPZformula} has been verified in many instances and has scored many successes. One of its major mysteries, however, is the famous $c=1$ barrier: $\g_{\rm str}$ becomes complex for central charges in the interval $1<c<25$. 

Very little is known about models of two dimensional gravity that go beyond the Liouville theory. However, it is very natural to seek for consistent generalizations, both from the physical and mathematical points of view. The Liouville action is special because it universally describes the coupling of any conformal field theory to gravity, but it says nothing about the much larger class of models obtained by coupling gravity to a non-conformal quantum field theory. Recently, it was shown in \cite{FKZMab} that, in an expansion in the inverse of the mass of a scalar matter field, the gravitational action picks a new term proportional to the so-called Mabuchi action. This result is quite interesting, because the Mabuchi action has many remarkable properties and has played a prominent r\^ole in the mathematical literature on K\"ahler geometry in recent years, see e.g.\ \cite{Mabmath}. The Mabuchi action satisfies the cocycle condition, which is a basic consistency requirement for any gravitational effective action. It is bounded from below and convex, which makes it a good candidate to be used in a path integral. Unlike the Liouville action, it is also defined in any complex dimension and thus provides a window on higher dimensional gravity as well. Finally, let us note that it was realized in \cite{K1} that the path integral involving the Mabuchi action appeared in disguise in \cite{ZW}, where it was proposed as an effective collective field theory for the droplets in the Quantum Hall regime. 


The goal of the present note is to continue a study, initiated in \cite{FKZMab,FKZ}, of the Mabuchi quantum gravity from first principles, with and without the presence of an additional Liouville term. In this context, it is very natural to parameterize the metrics of given area $A$ by their K\"ahler potential $\f$, instead of the more familiar conformal factor $\sigma$. The Mabuchi functional is most naturally expressed in terms of $\phi$ and the partition function at fixed area $Z_{\text{grav}}(A)$ is simply given by a path integral over $\phi$. Of course, the K\"ahler point of view ought to be equivalent to the description in terms of the conformal factor, since in two dimensions there is a one-to-one map between $(A,\phi)$ and $\sigma$.

A startling aspect of the K\"ahler point of view is that a natural non-perturbative regularization scheme exists \cite{FKZ}, in which the space of K\"ahler metrics is approximated by a finite dimensional space of Bergman metrics, which in turn are parameterized by Hermitian $N\times N$ matrices. The integer $N$ plays the r\^ole of a cutoff which must be sent to infinity. This approach to the quantum Mabuchi theory is pursued further in \cite{KZ}. In the present work we shall be more modest and use a simple perturbative approach. Our main goal is to generalize the formulae  \eqref{KPZformula}, \eqref{gstrexp} to include the effect of the Mabuchi action at one loop. Let us quote here our result for the one loop string susceptibility in the Liouville plus Mabuchi theory, $\beta^{2}$ being the coupling constant in front of the Mabuchi action (see Eq.\ \eqref{LpMaction}),
\be
\gamma_{\rm str}=\frac{h-1}{2}\k^2 -2\beta^2 +\frac{19-7h}{6} -\frac{4\b^2}{\k^2}\,\cdotp
\ee
A similar calculation was presented long ago by Zamolodchikov for the Liouville theory on the sphere \cite{zamolod}, much before the exact formula \eqref{KPZformula} was postulated. 
Unlike Zamolodchikov, who used the conformal factor and the $\zeta$-function regularization scheme, we shall use the K\"ahler formalism and a smooth spectral cutoff, along the lines developed in \cite{BF}.
This provides the simplest derivation of the one-loop partition functions and makes manifest the scheme-independence of the results. Moreover, this approach can be generalized to higher loops \cite{BF}; the two loop calculation will be presented elsewhere \cite{BF2loops}. We also discuss the $\zeta$-function regularization scheme, which introduces some interesting subtleties in the K\"ahler parameterization, related to the so-called multiplicative anomaly \cite{basicref,multianom}.

This paper is organized as follows. In the next section, we formulate and study two-dimensional quantum gravity in the K\"ahler formalism on an arbitrary Riemann surface of genus $h$, defining the integration measure over the space of metrics and the gravitational action which is an arbitrary combination of the Liouville and Mabuchi functionals. At one-loop, the corresponding partition function is given by a ratio of determinants of operators involving the Laplace operator. The regularization of these formal ratios of determinants is studied in Sec.\ 3, using the smooth spectral cutoff scheme. We show that the divergences can be absorbed in the standard cosmological constant counterterm, yielding a finite and scheme-independent string susceptibility. We also discuss the case of the pure Mabuchi theory, which is argued to correspond to a non-perturbative quantum gravity theory with a non-trivial UV fixed point. Finally, we conclude in Sec.\ 4. Another familiar regularization method is the $\zeta$-function scheme. In our case, this method introduces non-trivial subtleties associated with the multiplicative anomaly phenomenon. Unlike the known quantum field theoretic examples \cite{multianom}, where this anomaly is actually unphysical and can be absorbed in the local counterterms \cite{irrelevantMultianomaly}, it does play a non-trivial physical r\^ole in the present quantum gravity context. This is explained in details App.\ A where, in particular, we derive a new general multiplicative anomaly formula for Laplace-type operators on arbitrary compact Riemann surfaces. Of course, the correct analysis in the $\zeta$-function scheme reproduces the results obtained in the main text. We have also included a brief discussion of the sharp cutoff method for the round sphere in App.\ B.


\section{Liouville and Mabuchi quantum gravity}

\setcounter{equation}{0}

Our goal is to the study various partition functions of two-dimensional quantum gravity on compact Riemann surfaces of genus $h$ and in particular their dependence on the area $A$ of the surface. The basic ingredients are the correct definition of the measure on the space of metrics, as well as the relevant gravitational actions which can be the Liouville or Mabuchi actions, or a combination of both. While the Liouville action is usually written in terms of the conformal factor variable $\s$, the Mabuchi action is most naturally defined in terms of the K\"ahler potential $\f$. There is a one-to-one mapping between $\sigma$ and $(A,\phi)$.

\subsection{The measure on the space of metrics}\label{measureSec}

We pick a compact Riemann surface with fixed complex structure moduli. Modulo the action of diffeomorphisms, any metric $g$ on the surface can be written in the form
\be\label{confgauge} g = e^{2\sigma}g_{0}\, ,\ee
where $g_{0}$ is a reference metric which can be chosen to be the constant scalar curvature metric of some given area $A_{0}$. If $A$ is the area of the metric $g$, the K\"ahler potential is defined by 
\be\label{gg0sig}
e^{2\s}= \frac{A}{A_0}\left(1-\frac{1}{2} A_0 \D_0\f\right)\ .
\ee
Given $\sigma$, the above relation actually defines $A$ and $\phi$ uniquely, up to unphysical constant shifts of $\phi$. The relation \eqref{gg0sig} is equivalent to the relation
\be\label{kahlform} \omega =\frac{A}{A_{0}} \omega_{0} + i A\partial\bar\partial\phi\ee
between the volume (K\"ahler) forms of the metrics $g$ and $g_{0}$. For later convenience, we also introduce the constant scalar curvature metric $g_*$ of area $A$, with corresponding Laplace operator $\D_*$ and Ricci scalar $R_*$,
\be\label{gstar}
g_*=\frac{A}{A_0} g_0 \quad  ,\quad  \D_*=\frac{A_0}{A}\D_0 
\quad  ,\quad  R_*=\frac{A_0}{A} R_0 \ .
\ee 
 
Since we want to do quantum gravity, we will need to integrate over the space of metrics modulo diffeomorphisms. The integration measure on this space can be derived from a choice of metric on the space of metrics. It is customary to assume that the correct metric must be ultralocal and thus must take the form $||\dd g||^2=\int\d^2 x \sqrt{g}\ \dd g_{ab}\dd g_{cd}\, \big( g^{ac} g^{bd}+c\, g^{ab}g^{cd}\big)$ for some constant $c>-1/2$ (other choices may actually be natural and mathematically consistent \cite{FKZ}, see below). Using \eqref{confgauge}, this yields 
\be\label{sigmametric}
||\dd g||^2=8 (1+2c)  \ ||\dd\s||^2
\, ,\quad
||\dd\s||^2=
\int\d^2 x \sqrt{g_0}\ e^{2\s}\,(\dd\s)^2 \ .
\ee
The functional integration measure ${\cal D}\s$ over $\s$ is induced from this metric. Because of the non-trivial factor $e^{2\sigma}$, it is {\it not} the measure of a free field. Instead of $\sigma$, we can also use equivalently the variables $(A,\phi)$. These variables are particularly convenient when one works at fixed area. Using \eqref{gg0sig}, the metric \eqref{sigmametric} yields
\be\label{ddsigmaddphi}
||\dd\s||^2=\frac{(\dd A)^2}{ 4 A} + ||\dd\s||^2_A \ ,
\ee
with the $||\dd\s||^2_A$ being the metric on the space of metrics for fixed area $A$,
\be\label{sigmametric2}
||\dd\s||^2_A =\frac{1}{16} \int\d^2 x \sqrt{g}\, (A\D \dd\f)^2\, .\ee
In view of our later treatment of the semiclassical approximation, it is convenient to rewrite this metric as
\be\label{sigmametric3}||\dd\s||^2_A =\frac{1}{16}  \int\! \d^2 x\,\sqrt{g_0}\, e^{2\s} \Bigl( e^{-2\s} A \D_0\dd \f\Bigr)^2
\\
= \frac{1}{16} \int\!\d^2 x \sqrt{g_*}\, \bigl( 1-{\textstyle\frac{1}{2}} A\D_*\f \bigr)^{-1}\bigl( A\D_*\dd\f\bigr)^2
\ .\ee
Formally, \eqref{ddsigmaddphi} thus induces a measure
\be\label{measure2}\mathcal D\sigma = 
\frac{\d A}{\sqrt{A}}\,{\cal D}\phi =\frac{\d A}{ \sqrt{A}}\, \Bigl[\Det' \bigl( 1-{\textstyle\frac{1}{ 2}} A\D_*\f \bigr)^{-1}\Bigr]^{1/2} \Det' (A\D_*)\, {\cal D}_*\f \ ,
\ee
where ${\cal D}_*\f$ is the standard free field integration measure in the background metric $g_{*}$ deduced from the metric $||\dd\f||_{*}^2=\int\!\d^2 x \sqrt{g_*}\, \dd\f^2$ in the space of K\"ahler potentials. The notation $\Det'$ means that we are not taking into account the zero mode of the Laplacian when computing the determinant, consistently with the fact that the zero mode of $\phi$ is unphysical and thus must not be included in the integration measure over the K\"ahler potentials. The measure ${\cal D}_*\f$ can be expressed in the traditional way by expanding $\f$ in eigenmodes of the Laplace operator. If $0=\la_{0}<\la_{1}\leq\la_{2}\leq\cdots$ are the eigenvalues,
\be\label{eigenexp}
\f=\sum_{r>0} c_r \p_r\, , \quad
\D_* \p_r=\l_r\p_r\, , \quad
\int\!\d^2 x \sqrt{g_*} \,\p_r \p_s= \delta_{rs}\, ,
\ee
we have
\be\label{measureA}
{\cal D}_* \f=\prod_{r>0} \d c_r \ .
\ee
Of course, we could instead define an expansion of $\f$ with respect to eigenfunctions normalized with the metric $g_0$ of area $A_0$. Obviously,  one then has 
\be\label{lamcrrel}
\l_r=\frac{A}{A_0}\l_r^0 \, , \quad
c_r=\sqrt{\frac{A}{A_0}}\, c_r^0  \, , \quad
\p_r=\sqrt{\frac{A_0}{A}}\, \p_r^0
\ee
and the measure ${\cal D}_{0} \f=\prod_r \d c_r^0$ is related to ${\cal D}_* \f$ by ${\cal D}_* \f=e^{\frac{1}{2}\sum_r \ln\frac{A}{A_0}}\ {\cal D}_{0}\f$. Of course, all these relations are formal and must be regularized, as we shall discuss in the next section.

\smallskip

\noindent\emph{Remarks}

\noindent i) Writing the metric $g$ in the form \eqref{confgauge} amounts to fixing the action of the diffeomorphisms. This produces the well-known ghosts, whose effects can be absorbed in the coefficient $\kappa^{2}$ of the Liouville action defined in the next section (they contribute the 26 in \eqref{kappadef}). A further subtlety arises in the case of the sphere $h=0$, because the gauge-fixing \eqref{confgauge} is then incomplete. An additional gauge fixing of the residual $\text{SL}(2,\mathbb C)/\text{SU}(2)$ group of diffeomorphisms acting non-trivially on $\sigma$ and $\phi$ must be performed. The result of this procedure is simply to project out the spin-one modes of $\phi$ in the decomposition \eqref{eigenexp} and to produce an overall factor of $A^{3/2}$ in the partition function coming from the Faddeev-Popov determinant.

\noindent ii) The space of K\"ahler potentials $\phi$ over which we integrate with the measure \eqref{measure2} is really the set of functions expanded as in \eqref{eigenexp} constrained by the inequality
\be\label{phiconstraint}
A \D_* \f < 2 \ .\ee 
This condition comes from the positivity of the metric, see e.g.\ the formula \eqref{gg0sig}. In perturbation theory around $\phi=0$, \eqref{phiconstraint} is irrelevant and thus will not bother us in the present paper. However, it is crucial in cases where the perturbation theory breaks down or, more generally, in the non-perturbative definition of the integral over K\"ahler potentials. In \cite{FKZ}, it is explained how the constraint \eqref{phiconstraint} can be elegantly solved by using a matrix parameterization of the K\"ahler potentials, reducing in this way the non-perturbative quantum gravity path integral to a matrix model.

\subsection{Liouville and Mabuchi actions}

\subsubsection{Liouville action}

The Liouville action is given in terms of a fixed reference metric $g_0$ and the metric $g$ defined in \eqref{confgauge},
\be\label{Liouville}
S_{\text L}[g_{0},g]= \int\!\d^2 x \sqrt{g_0}\,\bigl( \s \D_0\s
+ R_0 \s\bigr) \, .
\ee
This action satisfies the so-called cocycle identity,
\be\label{cocyLiou} S_{\text L}[g,g'']=S_{\text L}[g,g']+S_{\text L}[g',g'']\, ,\ee
which is a fundamental consistency condition any gravitational action must satisfy (see e.g.\ \cite{FKZMab}). 

The semiclassical partition function at fixed area will be dominated by the minimum of the Liouville action at fixed area. The constraint of fixed area is most easily implemented by using the parameterization \eqref{gg0sig} in terms of $A$ and $\phi$ instead of $\sigma$. The variation of $S_{\text L}$ with respect to $\phi$ then yields
\be\label{Liouvvar}
\delta S_L=-\frac{A}{4}\int\!\d^2 x \sqrt{g}\, \Delta R[g]\, \delta\f\, .
\ee 
We thus find that the classical saddle point of the Liouville action is the metric of constant curvature, which is a well-known result. If we choose our reference metric $g_{0}$ to be of constant curvature, with area $A_{0}$, the saddle point value of $\sigma$ is $\sigma_{*}=\frac{1}{2}\ln\frac{A}{A_{0}}$ whereas $\phi_{*}$ is constant. Expanding around this saddle point up to quadratic order then leads to
\be\label{Liouvillefluc}
S_{\text L}[g_{0},g]=4\pi(1-h)\ln\frac{A}{A_{0}} + \frac{1}{16}\int\!\d^2 x\sqrt{g_*}\, \f\, (A\D_*)^2 \Bigl[\D_* +\frac{8\pi(h-1)}{A}\Bigr] \f
+O(\f^3) \, .\ee
The terms $O(\f^3)$ are relevant beyond the one loop approximation. Let us note that the zero modes of the operator governing the quadratic fluctuations are projected out; as stressed in \ref{measureSec}, the zero mode of $\phi$ is unphysical for all $h$ and, for $h=0$, the spin-one components, which are associated with three additional zero modes in this case, are eliminated by the full gauge-fixing of the group of diffeomorphisms.

\noindent\emph{Remark}: the full $A$-dependence of the Liouville action comes from the tree-level term $S_{\text L}(\sigma_{*})=4\pi(1-h)\ln (A/A_{0})$ in \eqref{Liouvillefluc}. The factors of $A$ in the quadratic fluctuations can be absorbed by using $A\Delta_{*}=A_{0}\Delta_{0}$ and $\sqrt{g_{*}} = \frac{A}{A_{0}}\sqrt{g_{0}}$. It is easy to check that this property actually remains true to all orders in the expansion. All the non-trivial area dependence of the quantum gravity partition function is thus a sort of anomaly and must come from the path integral measure or, more precisely, from the fact that a regularization procedure is required.

\subsubsection{Mabuchi action}

The Mabuchi action can be written in terms of the fixed reference metric $g_0$ and the metric $g$ defined in \eqref{confgauge} and parameterized as in \eqref{gg0sig},
\be\label{Mab1}
S_{\text M}[g_{0},g]=\int\d^2 x\, \sqrt{g_0} \left[ 2\pi(h-1)\f\D_0\f + \Bigl(\frac{8\pi(1-h)}{A_0}-R_0\Bigr) \f +\frac{4}{A} \s e^{2\s} \right] \, .\ee
This action is a well-defined functional of the metric $g$ because it is invariant under constant shifts of $\phi$, as can be easily checked by using the Gauss-Bonnet theorem. Morevoer, it satisfies the fundamental cocycle identity,
\be\label{cocyMab} S_{\text M}[g,g''] = S_{\text M}[g,g']+S_{\text M}[g',g'']\, ,\ee
showing that it is a priori a consistent gravitational action. This was explicitly demonstrated in \cite{FKZMab}, where it was shown that it enters when a scalar field with a small non-zero mass is coupled to gravity. 

Varying the metric $g$ at fixed area, or equivalently varying $\phi$, one gets
\be\label{Mabvar}
\delta S_{\text M}=\int\!\d^2 x\sqrt{g}\, \Bigl( \frac{8\pi(1-h)}{A}
-R\Bigr) \delta\f  \, ,\ee
which shows that the saddle point of $S_{\text M}$ is, as for the Liouville action, the metric of constant curvature. Expanding around this saddle point up to quadratic order, we then find
\be\label{Mabfluc}
S_{\text M}[g_{0},g]=2\ln\frac{A}{A_{0}} + \frac{1}{4}\int\!\d^2 x \sqrt{g_*}\, \f\, A\D_* \Bigl( \D_*+\frac{8\pi(h-1)}{A}\Bigr) \f + O(\f^3) \, .\ee
The operator governing the quadratic fluctuations is quite similar to the Liouville case, see \eqref{Liouvillefluc}. In particular, the zero modes are harmless because they have been projected out.

\noindent\emph{Remark}: as in the case of the Liouville action, the full $A$-dependence of the Mabuchi action comes from the tree-level term $2\ln(A/A_{0})$ in \eqref{Mabfluc}; this is true to all orders in the $\phi$-expansion.

\subsection{One-loop quantum gravity partition functions}\label{onelSec}

We want to compute the quantum gravity partition function with a total action which is a combination of both the Liouville and the Mabuchi functionals, which we write as
\be\label{LpMaction} S[g_{0},g] = \frac{\k^2}{8\pi} S_{\text L}[g_{0},g]+\b^2 S_{\text M}[g_{0},g]\, .\ee
The total partition function is given by
\be\label{QGZ}  Z_{\text{grav}} = \int\!\d A\, e^{-\mu^{2} A}Z(A;\kappa^{2},\beta^{2})\, ,\ee
where $\mu^{2}$ is an arbitrary cosmological constant and the partition function at fixed area is formally given in terms of the path integral measure $\mathcal D\phi$ defined in \eqref{measure2} by
\be\label{QGZ2} Z(A;\kappa^{2},\beta^{2}) =\frac{1}{\sqrt{A}} \int\!\mathcal D\phi\, e^{-\frac{\k^2}{8\pi\varepsilon} S_{\text L}[g_{0},g]-\frac{\b^2}{\varepsilon} S_{\text M}[g_{0},g]}\, .\ee
We have introduced a formal loop counting parameter $\varepsilon$ in terms of which one can write a loop expansion of the form
\be\label{loopExp} \ln Z = \sum_{L\geq 0}\varepsilon^{L-1}W_{L}\, .\ee
We shall limit ourselves to the one-loop approximation in the following. The tree-level contribution $W_{0}$ can be read from the tree-level terms in \eqref{Liouvillefluc} and \eqref{Mabfluc},
\be\label{W0result} W_{0} = \Bigl(\frac{h-1}{2}\kappa^{2}-2\beta^{2}\Bigr)\ln\frac{A}{A_{0}}\, \cdotp\ee
This yields a tree-level string susceptibility
\be\label{gammatree} \gamma_{\text{str}}^{\text{tree-level}} = \frac{h-1}{2}\kappa^{2} - 2\beta^{2}\ee
generalizing the leading term in \eqref{gstrexp}. At one-loop, the complicated determinant factor $\smash{\bigl[\Det' (1-\frac{1}{2}A\Delta_{*}\phi)^{-1}\bigr]^{1/2}}$ in the measure \eqref{measure2} is irrelevant and we can thus perform directly the Gaussian integration over $\phi$, using the expansions \eqref{Liouvillefluc} and \eqref{Mabfluc}, to get
\begin{multline}\label{QGZ3}
W_{1} = -\frac{1}{2}\ln A + \ln \Det' (A\D_*)\\
-\frac{1}{2} \ln \Det' \biggl[A\D_*
\Bigl( \D_*+\frac{8\pi(h-1)}{A}\Bigr)
\Bigl(\frac{\k^2}{32\pi}A\D_* +\b^2\Bigr) \biggr] \, , \ \text{for}\ h\ge 1\, .\end{multline}
On the sphere $h=0$, this formula must be slightly modified by removing the three spin-one zero modes from the determinant, which we indicate by the notation $\Det''$, and adding the associated contribution of $\frac{3}{2}\ln A$ from the Faddeev-Popov determinant, which yields
\be\label{QGZ3h0}
W_{1} = \ln A + \ln \Det'' (A\D_*)
-\frac{1}{2} \ln \Det'' \biggl[A\D_*
\Bigl( \D_*-\frac{8\pi}{A}\Bigr)
\Bigl(\frac{\k^2}{32\pi}A\D_* +\b^2\Bigr) \biggr] \, , \ \text{for}\ h=0\, .\ee
Of course, these expressions are formal and must be regularized, as discussed in the next section.

So far, we have been using the measure \eqref{measure2} to define the quantum gravity partition functions. This measure comes from the unique ultralocal metric on the space of metrics, as explained in \ref{measureSec}. In the context of quantum gravity, we believe that it is natural to consider the possibility that other measures could be relevant. Indeed, unlike in the case of local quantum field theories, the requirement of the ultralocality of the measure does not seem to have firm conceptual foundations. Instead of locality, the fundamental requirement in quantum gravity is rather  background independence and any measure that satisfies this constraint is a priori a good candidate to be used in the path integral. For example, a very natural background independent measure on the space of K\"ahler potentials is derived from the so-called Mabuchi metric
\be\label{Mabuchimet} ||\delta\phi||_{\text{M}}^{2} = \int\!\d^{2}x\sqrt{g}\,(\delta\phi)^{2}\, .\ee
The associated integration measure $\mathcal D_{\text{M}}\phi$ is formally related to the standard measure $\mathcal D\phi$ appearing in \eqref{measure2} by
\be\label{Mabuchimeasure}\mathcal D\phi =\mathcal D_{\text M}\phi\, \Det'(A\Delta)\, .\ee
Using this measure  $\mathcal D_{\text{M}}\phi$ , one can then define a ``Mabuchi'' quantum gravity path integral by
\be\label{MabZdef} Z_{\text M}(A;\kappa^{2},\beta^{2})=\int\!\mathcal D_{\text M}\phi\, e^{-\frac{\kappa^{2}}{8\pi}S_{\text L}[g_{0},g]-\beta^{2}S_{\text M}[g_{0},g]}\, .\ee
%


%
\section{Regularization}

In the previous section, we expressed the partition functions in terms of various determinants of Laplace-type operators or of products of such operators. These determinants are formal expressions that must  be regularized to obtain well-defined quantities.

A standard regularization and renormalization technique for determinants is the\break $\zeta$-function method. For a given operator $O$ with eigenvalues $o_n$, this  consists in defining the spectral $\zeta$-function as $\zeta_O(s)=\sum_n o_n^{-s}$  by analytic continuation from those values of $s$ where the sum converges absolutely, and defining the renormalized determinant by $\ln \Det_{(\zeta)} O =-\zeta'_O(0)$. However, this convenient and powerful method has several drawbacks. First, it does not allow a physical discussion of the divergences, as they are suppressed abstractly by the procedure of analytical continuation. Second, this method is essentially limited to one-loop computations. Third, one has to deal with the so-called multiplicative anomaly \cite{basicref,multianom}: the $\zeta$-renormalized determinant of a product of operators does not always equal the product of the $\zeta$-renormalized determinants of the operators. This ambiguity is usually harmless because it is associated with the familiar fact that the determinants are defined modulo the addition of arbitrary local counterterms. In this respect, the multiplicative ``anomaly'' is irrelevant \cite{irrelevantMultianomaly}. However, and quite remarkably, in the present quantum gravity context, this anomaly can have genuine physical effects. In particular, it is crucial to take into account the multiplicative anomaly when performing the change of variables from the conformal factor $\sigma$ to the area $A$ and K\"ahler potential $\phi$ in the path integral if one uses the $\zeta$-function regularization scheme. Overlooking this subtlety would yield the wrong answer for the area dependence of the partition function! All this is explained in details in Appendix A, which contains in particular the derivation of a general formula for the multiplicative anomaly in the $\zeta$ function scheme.

Another regularization scheme, which makes the discussion simpler and more physical, is the general spectral cutoff approach recently developed in \cite{BF}. As explained in details in \cite{BF}, this scheme can be viewed as a generalization of the $\zeta$-function approach while avoiding all the drawbacks just mentioned. We shall thus use this approach in the following.

At one-loop, the spectral cutoff scheme is extremely simple to describe because the generic one-loop quantity one has to compute takes the form of an infinite sum 
\be\label{formalsum} \sum_{r} G(\la_{r})\, ,\ee
where $G$ is a function of the eigenvalues $\l_r$ of an appropriate operator which for us is always the positive Laplacian $\Delta$. The spectral cutoff regularization uses a general cutoff function $f$, which is conveniently expressed as a Laplace transform
\be   f(x)=\int_0^\infty \d \a\, \vf(\a) e^{-\a x}\ee
and which must satisfy the condition
\be\label{fofzero} f(0) = 1\, .\ee
The formal sum \eqref{formalsum} is then defined as
\be\label{reg1}\sideset{}{^{f,\La,M}}
\sum_r G(\l_r)     = \sum_r f\Bigl(\frac{\l_r+M^2}{\L^2}\Bigr) G(\l_r)
=\int_0^\infty\! \d \a\, \vf(\a)
\Bigl[\sum_r G(\l_r)\Bigr]_{t=\frac{\alpha}{\L^2} ,\, M}\, ,\ee
where 
\be\label{bracknotdef}\Bigl[\sum_r G(\l_r)\Bigr]_{t,M}
=\sum_r \ e^{-t(\l_r+M^2)}\, G(\l_r)\,  ,\ee
$\L$ is the cutoff scale which is eventually sent to infinity and $M^2$ is an arbitrary fixed finite constant. Since $t=\a/\L^2$, the large $\L$ expansion of the regularized expression \eqref{reg1} is obtained from the small $t$ expansion of \eqref{bracknotdef}, see \cite{BF} for details.

Of course, even though all the regularized quantities depend on the cutoff function $f$ and the arbitrary parameter $M$, the physical quantities, which are defined up to the addition of local counterterms, must be independent of both $f$ and $M$. For the partition function \eqref{QGZ} or \eqref{QGZ2}, this means that all the $f$ and $M$ dependence can be absorbed in the cosmological constant $\mu^{2}$, possibly up to an area-independent global normalization constant.

\subsection{Regularized determinants}

Assume one wants to evaluate a general determinant of the form
\be\label{gendetform} \ln\Det\bigl( F_{1}(\Delta)\cdots F_{p}(\Delta)\bigr)\, ,\ee
where the operators $F_{i}(\Delta)$ can be expressed in terms of the Laplacian. Obviously, in the prescription \eqref{reg1}, the regularized version of \eqref{gendetform} is the sum of the regularized versions of the individual logarithm of determinants, simply because
\be\label{trivialidi} \sum_{r}f\Bigl(\frac{\l_r+M^2}{\L^2}\Bigr)
\ln\bigl(F_{1}(\la_{r})\cdots F_{p}(\la_{r})\bigr) = \sum_{i=1}^{p}
\sum_{r}f\Bigl(\frac{\l_r+M^2}{\L^2}\Bigr)\ln F_{i}(\la_{r})\, .
\ee
This shows that we do not have a multiplicative anomaly in our case.
Let us emphasize that this simple result is true because we use the same cutoff function $f$ and parameter $M$ to regularize all the sums and thus all the determinants. The multiplicative anomaly can only occur when one uses regularization schemes, like the $\zeta$-function scheme, for which this property is not true.

This being understood, we see that, to evaluate our one-loop quantum gravity partition function \eqref{QGZ3}, we simply need to compute the following two sums,
\begin{align}\label{calS0} {\cal S}_0 &=\sum_{r>0} e^{-t(\l_r+M^2)}\, ,\\
\label{calS1}
{\cal S}_1\bigl(a/A\bigr)&
=\sum_{r>0}  e^{-t(\l_r+M^2)}\, \ln ( \l_r + \frac{a}{A}) \ ,
\end{align}
which are such that
\be \label{detzzz} \Bigl[ \ln\Det' \Bigl(z\bigl(\Delta + a/A\bigr)\Bigr)\Bigr]_{t,M} = (\ln z)\mathcal S_{0}  + \mathcal S_{1}(a/A)\, .\ee
Following the techniques used in \cite{BF}, these sums are evaluated as follows\footnote{One first uses the identity $e^{-x}=\int_{b-i\infty}^{b+i\infty} \d s\, x^{-s}\G(s)$ valid for any $b>0$, as can be seen by closing the contour on a semi-infinite rectangle on the left in the complex plane and picking up all poles of $\G(s)$. The interchange of the sum and the integral is then justified if one chooses $b>1$ so that $\Re s>1$ and the sum converges absolutely since the large $r$ asymptotics of the $\l_r$ is governed by the Weyl law $\l_r\sim 4\pi r/A$.}
\begin{align}\nonumber\label{S1eval}
{\cal S}_1\Bigl(\frac{a}{A}\Bigr)&=e^{-t M^2+t\frac{a}{A}}\sum_{r>0} e^{-t(\l_r+\frac{a}{A})} \,\ln \left( \l_r +\frac{a}{A}\right)\\\nonumber &
=\frac{e^{-t M^2+t\frac{a}{A}}}{2\pi i} \int_{b-i\infty}^{b+i\infty} \d s\, \frac{\G(s)}{t^s}\,\sum_{r>0} \frac{\ln \left( \l_r +\frac{a}{A}\right)}{(\l_r +\frac{a}{A})^s}\\
&=-\frac{e^{-t M^2+t\frac{a}{A}}}{2\pi i}\int_{b-i\infty}^{b+i\infty} \d s\, \frac{\G(s)}{t^s} \zeta'_{\D}\bigl(s,\frac{a}{A}\bigr) \, ,
\end{align}
where we must choose $b>1$ and
\be\label{zetadef}
\zeta_{\D}\bigl(s,a/A\bigr)=\sum_{r>0} \frac{1}{(\l_r +\frac{a}{A})^s}
\ee 
is the spectral $\z$-function of the operator $\D+a/A$, excluding the zero mode. 
The integrand in \eqref{S1eval} has poles for all negative integer values of $s$, as well as at $s=0$ and $s=1$. As explained in \cite{BF}, the small $t$ asymptotic expansion of the integral can be evaluated by closing the contour with an infinite semi-rectangle on the left and picking up the contributions of all these poles. The poles at $s<0$ lead to contributions that are $O(t)$, i.e.\ $O(1/\L)$ and will vanish in the large $\L$-limit. Thus the only relevant contributions come from the poles at $s=0$ and $s=1$.
As is well known, the poles and residues of  $ \zeta_{\D}(s,\frac{a}{A})$, as well as the relevant finite values can be deduced from the small $\tau$ asymptotic expansion of the integrated heat-kernel $K(\tau)$ (see e.g.\ \cite{BF}, being careful that in the present case the zero mode is subtracted). One has
\begin{align}\label{zetapoles1}
\zeta_{\D}(0,\frac{a}{A})&=-\frac{h+2}{3}-\frac{a}{4\pi}\, \cvp
\\\label{zetapoles2}
 \zeta_{\D}(s,\frac{a}{A}) &\underset{s\rightarrow 1}{\sim} \frac{A}{4\pi} \, \frac{1}{s-1}\,\cdotp
\end{align}
Using also $\frac{\G(s)}{t^s}\sim
\frac{1}{s}$ as $s\to 0$ and $ \frac{\G(s)}{t^s}\sim \frac{1}{t} -\frac{\g+\ln t}{t}\, (s-1)$ as $s\to 1$, we obtain for the sum ${\cal S}_1$
\be\label{S1eval2}
{\cal S}_1\Bigl(\frac{a}{A}\Bigr)=\frac{A}{4\pi}\Bigl(-\frac{1}{t}+M^{2}\Bigr)\bigl(\gamma+\ln t\bigr)-\frac{a}{4\pi}\bigl(\gamma+\ln t\bigr)-\zeta'_{\Delta}\bigl(0,a/A\bigr) + O(t)\, .
\ee
The sum ${\cal S}_0$ is evaluated similarly and we find
\be\label{S0eval}
{\cal S}_0 =\frac{A}{4\pi}\Bigl(\frac{1}{t}-M^{2}\Bigr) -\frac{h+2}{3}+O(t)
\, .\ee

For our purposes, we shall need the area dependence of these expressions when the determinants are evaluated on the saddle point metric $g_{*}$ of area $A$. Using \eqref{gstar} we observe that the definition \eqref{zetadef} together with \eqref{lamcrrel} imply the scalings
\begin{align}\label{zetascaling1}
\zeta_{\D_*}\bigl(s,\frac{a}{A}\bigr)&=\Bigl(\frac{A}{A_0}\Bigr)^s \zeta_{\D_0}\bigl(s,\frac{a}{A_0}\bigr)\, ,\\\label{zetascaling2}
\zeta'_{\D_*}\bigl(0,\frac{a}{A}\bigr)&=-\Bigl(\frac{h+2}{3}+\frac{a}{4\pi}\Bigr) \ln\frac{A}{A_0}+ \zeta'_{\D_0}\bigl(0,\frac{a}{A_0}\bigr)\, ,
\end{align}
so that finally
\begin{multline}\label{S1eval3}
{\cal S}_1\Bigl(\frac{a}{A}\Bigr)
= \frac{A}{4\pi}\Bigl(-\frac{1}{t}+M^{2}\Bigr)\bigl(\gamma+\ln t\bigr)
+\Bigl(\frac{h+2}{3}+\frac{a}{4\pi}\Bigr) \ln\frac{A}{A_0}\\
-\frac{a}{4\pi}\bigl(\g+\ln t\bigr)
- \zeta'_{\D_0}\bigl(0,\frac{a}{A_0}\bigr) +O(t)\, .\end{multline}
Then, from \eqref{detzzz}, we get
\begin{multline}\label{zdet}
\Bigl[ \ln\Det' \Bigl(z\bigl(\Delta_{*} + a/A\bigr)\Bigr)\Bigr]_{t,M}=
\frac{A}{4\pi}\Bigl(\frac{1}{t}-M^{2}\Bigr)\bigl(\ln\frac{z}{t} - \gamma\bigr)\\+\Bigl(\frac{h+2}{3}+\frac{a}{4\pi}\Bigr) \ln\frac{A}{z\,A_0} + \frac{a}{4\pi}\bigl(\ln\frac{z}{t} - \gamma\bigr)
- \zeta'_{\D_0}\bigl(0,\frac{a}{A_0}\bigr) +O(t) \, .
\end{multline}
Note that upon plugging this result into a formula like \eqref{reg1}, 
the $t$-dependent terms on the right-hand side correspond to the divergent terms $\sim \L^2 \ln\L^2,\ \sim\L^2$ and $\sim \ln\L^2$.

\subsection{The renormalized quantum gravity partition functions} 

Using \eqref{zdet}, we immediately obtain the regularized version of the one-loop quantum gravity partition function \eqref{QGZ3} 
\begin{multline}\label{QGZ6}
\bigl[W_{1}\bigr]_{t,M}
= A\,\Bigl[\frac{1}{8\pi t} \Bigl(\g+\ln t-\ln\frac{\k^2}{32\pi} \Bigr) 
-\frac{M^2}{8\pi} \Bigl(\g+\ln t\Bigr)\Bigr]\\
-\Bigl[ \frac{1}{2} +\frac{7h-4}{6}+\frac{4\b^2}{\k^2}\Bigr] \ln\frac{A}{A_0} +C[A_0,\k^2,\b^2,t,M] \, ,
\end{multline}
where $C$ is an $A$-independent irrelevant coefficient. The first term in \eqref{QGZ6}, which is divergent and cutoff dependent, is proportional to $A$ and can thus be absorbed into the cosmological constant. The one-loop string susceptibility is read off from the coefficient of the $\ln A$ term,
\be\label{gammastring}
\g_{\rm str}^{\text{one-loop}}=\frac{19-7h}{6} -\frac{4\b^2}{\k^2}\,\cdotp
\ee
As required, it is finite and cutoff independent. Of course, it agrees with and generalizes \eqref{gstrexp}.

It remains to see how this is modified for genus 0. For $h=0$, we must start with \eqref{QGZ3h0}. The associated $\zeta$ functions are then defined excluding the spin-one modes. This yields an additional $-\frac{3}{2}\ln A$ contribution to the partition function which cancels the $+\frac{3}{2}\ln A$ term discussed in \ref{onelSec} that came from the Faddeev-Popov determinant. Overall, the formula \eqref{gammastring} is also valid at $h=0$.

\smallskip

\noindent\emph{Remark}: Let us make some brief remarks about the higher loop contributions to the partition function. They come from vacuum diagrams involving one or more vertices. The latter are, of course, determined by the terms of order $\f^n,\ n\ge 3$ in the expansions \eqref{Liouvillefluc} and \eqref{Mabfluc} of the Liouville and Mabuchi actions. Note that a term of order $\f^n$ in the Liouville action also involves a factor $(A\D_*)^n \D_*$ or $(A\D_*)^n$, and a term of order $\f^n$ in the Mabuchi action is accompanied by a factor $(A\D_*)^{n-1}\D_*$. One should also keep in mind that the propagator that follows from \eqref{LpMaction} is the inverse of
\smash{$A\D_* \bigl( A\D_* +\frac{4\b^2}{\k^2}\bigr)\bigl(\D_*+\frac{8\pi(h-1)}{A}\bigr)$}.
In addition, one also has to take into account the contributions from the non-trivial determinant in \eqref{measure2}. Standard power-counting then shows that all these interaction vertices are renormalizable: the superficial degree of divergence of any vacuum loop diagram one can make from these vertices and the propagator is at most two. We will report on the two-loop contribution in \cite{BF2loops}.


\subsection{On the pure Mabuchi quantum gravity}

It is also interesting to study a two-dimensional quantum gravity involving only the Mabuchi action. It was shown in \cite{FKZMab} that this case can occur when coupling a two dimensional QFT to gravity. However, the semi-classical analysis of the pure Mabuchi partition function is non-trivial. At tree-level, the result \eqref{W0result} for $\kappa=0$ is of course valid, but at one-loop we cannot use \eqref{QGZ6}. The problem is manifest in \eqref{LpMaction}, where the operator $\frac{\k^2}{32\pi}A\D_* +\b^2$ appears, which shows that setting the Liouville coefficient $\k^2$ to zero completely changes the UV behaviour. It is actually very simple to check that, contrary to the combined Liouville plus Mabuchi theory, the pure Mabuchi model is not perturbatively renormalizable, any $L$-loop vacuum diagram having a superficial degree of divergence $2L$. Of course the same conclusion is reached for a model with non-zero but fixed $\kappa$, in the limit $\beta\rightarrow\infty$. 

It is nevertheless quite instructive to work out explicitly what happens at one-loop. Choosing $h\geq 1$ and starting with \eqref{QGZ3} at $\k^2=0$,
\be\label{QGZMab1}
W_{1} = -\frac{1}{2}\ln A + \ln \Det' (A\D_*)
-\frac{1}{2} \ln \Det' \biggl[\b^2\,A\D_*
\Bigl( \D_*+\frac{8\pi(h-1)}{A}\Bigr)
\biggr] \, ,\ee
we get, using \eqref{zdet},
\be\label{QGZMab2}
\bigl[W_{1}\bigr]_{t,M}
=  - \frac{7h-1}{6}\ln\frac{A}{A_0} 
-\frac{A}{8\pi}\Big(\frac{1}{t}-M^2\Big) \ln\frac{\b^2}{4\pi A} +C_{\rm Mab}[A_0,\b^2,t,M] \ ,
\ee
with some $A$-independent coefficient $C_{\rm Mab}$. This formula shows that a new $\La^{2} A\ln A$ divergence appears, which cannot be absorbed in a standard local counterterm. The same problem occurs for the pure Mubuchi path integral \eqref{MabZdef} defined with the Mabuchi metric \eqref{Mabuchimet}.

It is interesting to note that these problems can be avoided, at one-loop, if one uses a rescaled Mabuchi metric of the form
\be\label{newMab} ||\delta\phi||^{2}_{\text{M'}} = A\int\!\d^{2}x\sqrt{g}\, (\delta\phi)^{2}\, .\ee
With this metric, the one-loop contribution reads, up to area-independent terms,
\be\label{MabQGZ}
W_{1}= -\frac{1}{2}\ln \Det'\left[\D_*\Bigl(\D_*+\frac{8\pi(h-1)}{A}\Bigr)\right] \ .
\ee
Upon regularizing as before, this yields
\be\label{MabQGZ3}
\bigl[W_{1}\bigr]_{t,M} =
-\frac{4h-1}{3}\ln\frac{A}{A_0} 
+\frac{A}{4\pi }\Big(\frac{1}{t}-M^2\Big)\left(\g+\ln t\right) 
+ \wh C_{\rm Mab}[A_0,t,M] \, ,
\ee
with some constant $\wh C_{\rm Mab}$. This result is consistent, in the sense that it does not have divergences proportional to $A\ln A$. It yields
\be\label{pureMga} \gamma_{\text{str}}^{\text{one-loop}} = \frac{10-4h}{3}\ee
for the one-loop string susceptibility of the pure Mabuchi theory defined with the metric \eqref{newMab}.

Of course, at higher loop orders, we run into the problem of the non-renor\-ma\-li\-za\-bi\-lity of the model. This non-renormalizability comes from interaction vertices of the form $(\Delta\phi)^{n}$ in the Mabuchi action which diverge in the UV. In other words, the model becomes strongly coupled in the UV and cannot be described by perturbing a Gaussian fixed point. However, this does not imply that the model is not well-defined at the non-perturbative level. The UV behaviour can,  in principle, be governed by a non-trivial UV fixed point which  provides a non-perturbative definition. Actually, we believe that this possibility is very plausible in the case of the Mabuchi theory. Indeed, the growth of the interactions $(\Delta\phi)^{n}$ is tamed in the full, non-perturbative, model, thank's to the fundamental inequality \eqref{phiconstraint}. This constraint is na\"\i vely irrelevant in perturbation theory, but the non-renormalizability implies that it will actually always play a crucial r\^ole, modifying drastically the UV behaviour of the theory. A promising approach to handle the non-perturbative Mabuchi model is then to use the formalism developed in \cite{FKZ}, which is based on a parameterization of the metrics which implement automatically the constraint \eqref{phiconstraint}. First steps in this direction will be presented in \cite{KZ}.


\section{Conclusion}

We have studied various two-dimensional quantum gravity partition functions in the path integral approach with various integration measures, the gravitational action being a combination of the Liouville and Mabuchi functionals. At one loop, the partition functions are given by ratios of determinants of Laplace-type operators on general Riemann surfaces. These determinants are most easily computed using a smooth spectral cutoff regularization. We have obtained in this way the regulator-independent string susceptibility $\gamma_{\text{str}}$ at one-loop, generalizing the famous result of the Liouville theory. The extension of this calculation to the two-loop order will be reported in a forthcoming publication \cite{BF2loops}, using the multiloop spectral cutoff formalism developed in \cite{BF}.

We believe that the generalization of the standard Liouville model by the Mabuchi term yields an extremely natural and interesting theory, both from the physical and the mathematical points of view. It opens interesting new directions of research in the old and venerable field of two-dimensional quantum gravity and random surfaces. In particular, it would be extremely interesting to see if a general reasoning, based on the background independence of the theory, would allow to find the exact string susceptibility with the Mabuchi term, generalizing the celebrated KPZ result \eqref{KPZformula}. The Mabuchi model also opens a window on higher dimensional gravity theory \cite{KZ}, since the Mabuchi action, unlike the Liouville action, admits natural generalization in any complex dimensions, a context in which it has been much studied in the mathematical literature. 

\subsection*{Acknowledgments}

We would like to thank Steve Zelditch for useful discussions on this and related matters. The work of F.F.\ is supported in part by the belgian FRFC (grant 2.4655.07) and IISN (grant 4.4511.06 and 4.4514.08).
S.K.\ is supported by the postdoctoral fellowship from the Alexander von Humboldt Foundation, by the grants DFG-grant ZI 513/2-1, RFBR 12-01-00482, RFBR 12-01-33071 (mol$\_$a$\_$ved), NSh-3349.2012.2, and by Ministry of Education and Science of the Russian Federation under the contract 8207.

\begin{appendix}

\section{Multiplicative anomaly and $\zeta$-function}

\subsection{Generalities}

Let $L_{1},\ldots,L_{n}$ be self-adjoint operators acting on a Hilbert space $\mathscr H$. The multiplicative anomaly $a (L_{1},\ldots,L_{n})$ is defined by the equation
\be\label{anomdef}e^{a (L_{1},\ldots,L_{n})} =
\frac{\Det\prod_{i=1}^{n}L_{i}}{\prod_{i=1}^{n}\Det L_{i}}\,\cdotp\ee
In finite dimension, the determinant of a product of linear operators is the product of the determinants and thus the multiplicative anomaly automatically vanishes. In infinite dimension, however, the definition of the determinants requires a suitable renormalization procedure that can violate this simple property of finite-dimensional determinants. The multiplicative anomaly \eqref{anomdef} can then be non-trivial \cite{basicref}. It has been studied in a number of special cases involving the product of two Laplace-type operators, using the $\zeta$-function regularization method and the Wodzicki residue formula \cite{multianom}. For example, we shall derive below, in the $\zeta$-function scheme where 
\be\label{zetadetdef}
\Det_{(\zeta)} L = \exp(-\zeta_L'(0))\ ,
\ee
a new generalized multiplicative anomaly formula of the form
\be\label{anomdef2}\Det_{(\zeta)}\biggl[\frac{\prod_{i=1}^{n}L_{i}}{\prod_{j=1}^{m}\tilde L_{j}}\biggr] = e^{a_{(\zeta)}(L_{1},\ldots,L_{n};\tilde L_{1},\ldots,\tilde L_{m})}\frac{\prod_{i=1}^{n}\Det_{(\zeta)} L_{i}}{\prod_{j=1}^{m}\Det_{(\zeta)}\tilde L_{j}}\,\cvp \quad m\not = n,\ee
for shifted Laplace operators 
\be\label{LtildeL} L_{i}=\Delta_{i}+ a_{i}/A_{i}\, ,\quad\tilde L_{j}=
\tilde\Delta_{j}+\tilde a_{j}/\tilde A_{j}\, ,\ee
where 
\be\label{deltaoprel} \Delta_{i}=\frac{A}{A_{i}}\Delta\, ,\quad
\tilde\Delta_{j}=\frac{A}{\tilde A_{j}}\Delta\ee
are the Laplacians for metrics $g_{i}=(A_{i}/A)g$ and $\tilde g_{j}=(\tilde A_{j}/A)g$ of areas $A_{i}$ and $\tilde A_{j}$ respectively and where $a_{i}$ and $\tilde a_{j}$ are real dimensionless constants. We shall show that
\begin{multline}\label{anomaly} a_{(\zeta)}\bigl(L_{1},\ldots,L_{n};\tilde L_{1},\ldots,\tilde L_{m}\bigr) =\\\frac{1}{4\pi(n-m)}\Biggl[ \sum_{i=1}^{n}\biggl(
\sum_{k=1}^{n}a_{k}-\sum_{k=1}^{m}\tilde a_{k}-(n-m)a_{i}\biggr)\ln A_{i}
\\-\sum_{j=1}^{m}\biggl(
\sum_{k=1}^{n}a_{k}-\sum_{k=1}^{m}\tilde a_{k}-(n-m)\tilde a_{j}\biggr)\ln\tilde A_{j}\Biggr]\, .
\end{multline}

To our knowledge, in all the examples that have been studied so far, the multiplicative anomaly does not have any physical effect \cite{irrelevantMultianomaly}. It is actually not an anomaly in the usual field theoretic meaning of the term, because it can be absorbed in local counterterms. Typically, the multiplicative anomaly is indeed proportional to the volume of space-time and its inclusion simply amounts to redefining the cosmological constant. However, we are going to explain below that in the context of the present paper, we come across a problem where overlooking the multiplicative anomaly would simply yield the wrong physical answer. This is related to the fact that the masses appearing in the shifted Laplace operators we have to deal with, as in  \eqref{LtildeL}, are actually inversely proportional to the areas.

\subsection{Application to change of variables in the path integral}

An interesting consequence of the multiplicative anomaly is as follows.
Let us consider the path integral
\be\label{pathintex1} \mathcal I=\int\!\mathcal D\varphi\, e^{-\frac{1}{2}\varphi\cdot M\cdot\varphi}\ee
over a scalar field $\varphi$, for some positive-definite symmetric operator $M$, and let us make the linear change of variables
\be\label{covar} \varphi = L\cdot\phi\ee
for some other positive-definite symmetric operator $L$. The path integral \eqref{pathintex1} can then be equivalently expressed as
\be\label{pathintex2} \mathcal I=\int\!\mathcal D\phi\, J e^{-\frac{1}{2}\phi\cdot LML\cdot\phi}\, ,\ee
for some Jacobian factor $J$ associated with the change of variable \eqref{covar}. Using the well-known result for gaussian path integrals, the equality between \eqref{pathintex1} and \eqref{pathintex2} is achieved by choosing
\be\label{Dphitrans} J =
\sqrt{\frac{\Det (LML)}{\Det M}} = e^{\frac{1}{2} a(L,M,L)}\Det L\, ,
\ee
where the multiplicative anomaly $a(L,M,L)$ is defined as in \eqref{anomdef}. \emph{This shows that the multiplicative anomaly modifies the classical formula for the transformation of the path integral measure under a linear change of variables of the form \eqref{covar}.}

Let us now illustrate this result, in the case of the pure Liouville theory for $h\geq 1$ for simplicity, for the change of variables from the conformal factor $\sigma$ of the metric defined by \eqref{confgauge} to the K\"ahler variables $(A,\phi)$ defined by \eqref{gg0sig}. In the $\zeta$-function scheme, if one uses the variable $\sigma$, the one-loop contribution to $\ln Z$ is simply found to be
\be\label{ZLsigmavar} W_{1} = -\frac{1}{2}\ln A -\frac{1}{2}\ln\Det_{(\zeta)}' \Bigl(\Delta_{*}+\frac{8\pi(h-1)}{A}\Bigr)\, ,\ee
whereas, if one uses the variable $\phi$, one finds
\begin{multline}\label{ZLphivar} W_{1} = -\frac{1}{2}\ln A+\ln\Det_{(\zeta)}'\bigl(A\Delta_{*}\bigr) -\frac{1}{2}\ln\Det_{(\zeta)}'\biggl[\bigl(A\Delta_{*}\bigr)^{2} \Bigl(\Delta_{*}+\frac{8\pi(h-1)}{A}\Bigr)\biggr]\\ + \frac{1}{2}a_{(\zeta)}\bigl(A\Delta_{*},A\Delta_{*},\Delta_{*}+8\pi(h-1)/A\bigr)
\, .\end{multline}
This formula is similar to \eqref{QGZ3}, but it now takes into account the multiplicative anomaly, which is non-trivial in the $\zeta$-function scheme but was absent in the regularization scheme used in the main text. Using the properties of the $\zeta$-regularized determinants that follow from \eqref{zdep} and \eqref{zetascaling2}, one immediately finds that the $A$-dependent piece in $W_1$ as given by \eqref{ZLsigmavar} is 
\be\label{W1sig} W_{1} = \frac{1-7h}{6}\ln A\ee
while the $A$-dependent piece in $W_1$ as given by \eqref{ZLphivar} is
\be\label{W1phi} W_{1} = -\frac{1+h}{2}\ln A + \frac{1}{2} a\bigl(A\Delta_{*},A\Delta_{*},\Delta_{*}+8\pi(h-1)/A\bigr) \ .
\ee
On the other hand, we can compute the multiplicative anomaly from \eqref{anomaly}, or  alternatively from applying \eqref{zdep} twice, with the result
\be\label{anomhere} a\bigl(A\Delta_{*},A\Delta_{*},\Delta_{*}+8\pi(h-1)/A\bigr) = -\frac{4(h-1)}{3}\ln A\, .\ee
We see that, by taking into account this anomaly, \eqref{W1phi} and \eqref{W1sig} agree with each other, and with  \eqref{QGZ6} obtained by using the spectral cutoff in the main text.

A compact way to present the above result is simply to include the correct Jacobian factor \eqref{Dphitrans} in the change of variables from $\sigma$ to $(A,\phi)$. The relation $\mathcal D\sigma = \frac{\d A}{\sqrt{A}}\Det'(A\Delta)\mathcal D_{\text M}\phi$ used in the main text in the spectral cutoff scheme (see \eqref{measure2} and \eqref{Mabuchimeasure}) must be replaced by
\be\label{zetameasrel} \bigl(\mathcal D\sigma\bigr)_{(\zeta)} =A^{\frac{1-4h}{6}} \d A\, \Det_{(\zeta)}'(A\Delta)\bigl( \mathcal D_{\text M}\phi\bigr)_{(\zeta)}\ee
in the one-loop $\zeta$ function scheme. The non-trivial factor of $A$ in this formula already appeared in \cite{FKZ} in the case of the sphere, $h=0$.

\subsection{Proof of the multiplicative anomaly formula}

Let us start by proving the

\noindent\textsc{Proposition}: \emph{The infinite dimensional determinants being defined using the $\zeta$~function regularization procedure, we have, for $n\ne m$,}
\be\label{zdep} \Det' _{\zeta}\biggl[z \frac{
\prod_{i=1}^{n}\bigl( \Delta + a_{i}/A \bigr)}{
\prod_{i=1}^{m}\bigl( \Delta + b_{i}/A \bigr)}
\biggr] = z^{-\frac{h+2}{3}-\frac{1}{4\pi (n-m)}(\sum_{i=1}^{n}a_{i}
-\sum_{i=1}^{m}b_{i})}\
\frac{
\prod_{i=1}^{n}\Det' _{\zeta}\bigl( \Delta + a_{i}/A \bigr)}{
\prod_{i=1}^{m}\Det' _{\zeta}\bigl( \Delta + b_{i}/A \bigr)}\, \cdotp
\ee
In particular, for $n=1$ and $m=0$, one simply has
\be\label{lemma1} \Det' _{\zeta}\Bigl( z\bigl(\Delta + a/A\bigr)\Bigr) = z^{-\frac{h+2}{3} - \frac{a}{4\pi}}\
\Det' _{\zeta} \bigl(\Delta + a/A\bigr)\, ,\ee
while for $z=1$ there is no multiplicative anomaly:
\be\label{dep} \Det' _{\zeta}\biggl[\frac{
\prod_{i=1}^{n}\bigl( \Delta + a_{i}/A \bigr)}{
\prod_{i=1}^{m}\bigl( \Delta + b_{i}/A \bigr)}
\biggr] = 
\frac{
\prod_{i=1}^{n}\Det' _{\zeta}\bigl( \Delta + a_{i}/A \bigr)}{
\prod_{i=1}^{m}\Det' _{\zeta}\bigl( \Delta + b_{i}/A \bigr)}\, \cdotp
\ee

We will first prove \eqref{lemma1} and \eqref{dep},  and then show that they imply \eqref{zdep}. Obviously, \eqref{lemma1} follows immediately from \eqref{zetascaling2} upon setting $A_0=A/z$. To prove \eqref{dep} we will need to study the  $\zeta$-function $\zeta(s,O(a_i,b_i))\equiv\zeta_{O(a_i,b_i)}(s)$ associated with the operator 
\be\label{Odef}
O(a_i,b_i)\equiv \frac{\prod_{i=1}^{n}(\Delta + a_{i}/A)}{\prod_{i=1}^{m}
(\Delta + b_{i}/A)} 
\ee
and express it in terms of the $\zeta$-functions $\zeta(s;\Delta + a/A)$ associated with the operators of the form $\Delta + a/A$. This can be done by using the following identity, valid for $n\ne m$,
\begin{multline}\label{ident} 
\frac{\prod_{i=1}^{m}(\lambda + b_{i}/A)^{s}}{\prod_{i=1}^{n}(\lambda + a_{i}/A)^{s}} =\frac{\Gamma\bigl((n-m)s\bigr)}{\Gamma(s)^{n}\Gamma(-s)^{m}}\int_{[0,1]^{n+m}}\!\!\!\d x_{1}\cdots\d x_{n+m}\, \delta\Bigl(\sum_{i=1}^{n+m}x_{i}-1\Bigr)
\\
\times\ \frac{x_{1}^{s-1}\cdots x_{n}^{s-1}x_{n+1}^{-s-1}\cdots x_{n+m}^{-s-1}}{\bigl(\lambda + \sum_{i=1}^{n}a_{i}x_{i}/A + 
\sum_{i=1}^{m}b_{i}x_{n+i}/A\bigr)^{(n-m)s}}\,\cvp
\end{multline}
from which follows the relation
\begin{multline}\label{ident2} 
\zeta(s;O(a_i,b_i)) =
\frac{\Gamma\bigl((n-m)s\bigr)}{\Gamma(s)^{n}\Gamma(-s)^{m}}
\int_{[0,1]^{n+m}}\!\!\!\d x_{1}\cdots\d x_{n+m}\, \delta\Bigl(\sum_{i=1}^{n+m}x_{i}-1\Bigr)
\\
\times\ x_{1}^{s-1}\cdots x_{n}^{s-1}x_{n+1}^{-s-1}\cdots x_{n+m}^{-s-1}\,
\zeta\left((n-m)s; \D+\textstyle{\frac{\sum_{i=1}^{n}a_{i}x_{i}+\sum_{i=1}^{m}b_{i}x_{n+i}}{A}}\right)\, .
\end{multline}
It is straightforward to take the $s\to 0$ limit on both sides of (\ref{ident2}), using in particular (\ref{zetapoles1}) with $\wh a=\sum_{i=1}^{n}a_{i}x_{i} +
\sum_{i=1}^{m}b_{i}x_{n+i}$
and performing the resulting integrals. We obtain in this way
\be\label{zetaatzerogen} \zeta\bigl(0;O(a_i,b_i)\bigr)
= -\frac{h+2}{3}
-\frac{1}{4\pi (n-m)}\Bigl(\sum_{i=1}^{n}a_{i} -
\sum_{i=1}^{m}b_{i}\Bigr)\, .
\ee
In order to also obtain the terms of order $s$ in (\ref{ident2}), we  expand both sides of this equation up to terms  $O(s^{2})$. Using
$\zeta(s;O(a_i,b_i)) =
\zeta(0;O(a_i,b_i)) + s\,\zeta'(0;O(a_i,b_i)) + \mathcal O(s^{2})$
and
$\zeta\left((n-m)s; \D+\textstyle{\frac{\wh a}{A}}\right) 
= \zeta\left(0; \D+\textstyle{\frac{\wh a}{A}}\right) 
+ (n-m) s\ \zeta'\left(0; \D+\textstyle{\frac{\wh a}{A}}\right) +  O(s^{2})$
in (\ref{ident2}) yields
\begin{multline}\label{calczprime} 
\zeta'(0;O(a_i,b_i)) =
\lim_{s\rightarrow 0} (-1)^{m}s^{n+m-1}
\int_{[0,1]^{n+m}}\!\!\!\d x_{1}\cdots\d x_{n+m}\, \delta\Bigl(\sum_{i=1}^{n+m}x_{i}-1\Bigr)
\\
\times \ x_{1}^{s-1}\cdots x_{n}^{s-1}x_{n+1}^{-s-1}\cdots x_{n+m}^{-s-1}
\,\zeta'\left(0; \D+\textstyle{\frac{\sum_{i=1}^{n}a_{i}x_{i}+\sum_{i=1}^{m}b_{i}x_{n+i}}{A}}\right) \, .
\end{multline}
In the limit $s\rightarrow 0$, we pick only the leading pole in the integral (\ref{calczprime}). This pole comes from the integration regions in which all the variables $x_{i}$ except one are very near zero. For example, the contribution from the region in which 
$$(x_{2},\ldots,x_{n+m})\in[0,\epsilon]^{n+m-1}\, ,$$
with a very small $\epsilon$ and thus $x_{1}\simeq 1$ due to the constraint $\sum_{i}x_{i}=1$, is given by
\be\label{intcontr} 
\int_{[0,\epsilon]^{n+m-1}}\!\!\!\d x_{2}\cdots\d x_{n+m}\,
x_{2}^{s-1}\cdots x_{n}^{s-1}x_{n+1}^{-s-1}\cdots x_{n+m}^{-s-1}\,\zeta'(0,a_{1})
\underset{s\rightarrow 0}{\sim}\frac{\zeta'(0,\D+a_{1}/A)} {s^{n-1} (-s)^m}\, \cdotp
\ee
Regions in which $x_{i}\simeq 1$ for $2\leq i\leq n+m$ contribute in a similar way. Summing up all these contributions, (\ref{calczprime}) then yields
\be\label{finalzetasum} 
\zeta'\bigl(0;O(a_i,b_i)\bigr)= \sum_{i=1}^{n} \zeta'(0,\D+a_{i}/A)-\sum_{i=1}^{m}\zeta'(0,\D+b_{i}/A)\, ,\ee
which is equivalent to the equation (\ref{dep}) that we wanted to prove.

To see that \eqref{lemma1} and \eqref{dep} imply \eqref{zdep}, we first write
\be\label{st1}
\Det'_{\zeta}\biggl[z\frac{\prod_{i=1}^{n}\bigl( \Delta + a_{i}/A
\bigr)}{\prod_{i=1}^{m}\bigl( \Delta + b_{i}/A
\bigr)} \biggr] = \Det'_{\zeta}\biggl[\frac{\prod_{i=1}^{n}\bigl( \wh\Delta + a_{i}/\wh A
\bigr)}{\prod_{i=1}^{m}\bigl( \wh\Delta + b_{i}/\wh A
\bigr)} \biggr]\, ,\ee
where $\wh\Delta$ and $\wh A$ are the Laplacian and area associated with the rescaled metric 
\be\label{rescale2}
\wh g = z^{-\frac{1}{n-m}}g
\quad\Rightarrow\quad
\wh A = z^{-\frac{1}{n-m}}\, A \ ,\quad
\wh\D= z^{\frac{1}{n-m}} \D\, .
\ee
We then use (\ref{dep}), written for $\wh\D$ and $\wh A$ and finally \eqref{lemma1} in the form
\be\label{st3}
\Det'_{\zeta}\bigl(\wh\Delta + a_{i}/\wh A\bigr) = \Det'_{\zeta}\Bigl(
z^{\frac{1}{n-m}}\bigl(\Delta + a_{i}/ A\bigr)\Bigr) = 
z^{-\frac{h+2}{3(n-m)} - 
\frac{a_{i}}{4\pi (n-m)}} \Det'_{\zeta}\bigl(\Delta + a_{i}/A\bigr)\, ,
\ee
and similarly for $\Det'_{\zeta}(\wh\Delta + b_{i}/\wh A)$, to get \eqref{zdep}. 

The general anomaly formula \eqref{anomaly} then follows straightforwardly from \eqref{zdep} by using suitable rescalings, similar to Eq.\ \eqref{st1}, to put the determinants in the form of \eqref{zdep}.

\section{Sharp spectral cutoff on the sphere}
\setcounter{equation}{0}

Instead of the general smooth spectral cutoff used in the main text, one might want to try a sharp (hard) spectral cutoff instead. As extensively discussed in \cite{BF}, hard cutoff methods are plagued with many difficulties which make them inconsistent in general cases. However, we found it instructive to try the method on the very special case of the round sphere, for which one can make sense of it. One motivation for this computation stems from the matrix approach of \cite{FKZ} which is akin to a 
sharp cutoff method.

We thus consider the standard unit radius round sphere. The eigenvalues of the Laplacian are $l(l+1)$, $l\geq 0$, and the eigenfunctions are the spherical harmonics $Y_l^m$, $-l\leq m\leq l$. For the round sphere of area $A$, the eigenvalues are
\be\label{spherelam}
\l_{l,m}=\frac{4\pi}{A}\, l(l+1) \ .
\ee
The sharp cutoff consists in excluding all eigenvalues with $l> N$ for some large $N$. The regularized sums  ${\cal S}_0$ and ${\cal S}_1$ now are replaced by
\be\label{Tdef}
{\cal T}_0=\sum_{l=1}^N \sum_{m=-l}^l 1\, ,
\qquad 
{\cal T}_1\Big(\frac{a}{A}\Big)=\sum_{l=1}^N \sum_{m=-l}^l \ln\Bigl(\l_{l,m}+\frac{a}{A}\Bigr) \, .
\ee
where the zero-mode $l=0$ has again been excluded. With this sharp cutoff, the basic determinant \eqref{zdet} becomes
\be\label{zdetN}
\Bigl[\ln\Det'\Bigl( z \bigl(\D_*+a/A\bigr)\Bigr)\Bigr]_N=(\ln z)\, {\cal T}_0 
+ {\cal T}_1\Big(\frac{a}{A}\Big) \ ,
\ee
Obviously,
\be\label{T0}
{\cal T}_0=(N+1)^2-1
\ee
just counts the number of eigenvalues included in the sum, while
\be\label{T1}
{\cal T}_1\Big(\frac{a}{A}\Big)=\sum_{l=1}^N (2 l+1) \ln \left( l(l+1)+\frac{a}{4\pi}\right)
+\ln\frac{4\pi}{A}\ {\cal T}_0 \ .
\ee
This sum can be evaluated using the Euler-MacLaurin formula\footnote{
It reads $\sum_{l_0}^N f(l)=\int_{l_0}^N \d x\, f(x) +\frac{1}{2}\big(f(N)+f(l_0)\big)+\frac{1}{12}\big(f'(N)-f'(l_0)\big)-\frac{1}{180}\big( f'''(N)-f'''(l_0)\big)+\ldots$ with $f(l)=(2l+1)\ln\left( l(l+1)+\frac{a}{A}\right)$.
}
which yields
\begin{multline}\label{T1-2}
{\cal T}_1\Big(\frac{a}{A}\Big)=N(N+2)\ln\frac{4\pi}{A}
+\Bigl[N(N+2)+\frac{2}{3}+\frac{a}{4\pi}\Bigr] \ln\Bigl(N(N+1)+\frac{a}{4\pi}\Bigr)
\\-N(N+1)
+ c[a] +O\Bigl(\frac{1}{N}\Bigr)\, ,
\end{multline}
where $c[a]$ is some constant that does not depend on $N$ or $A$.

The next step is to identify $N$ in some way with the physical cutoff scale $\L$. With  the smooth cutoff, the spectrum is effectively cut off if $\l_r$ is (much) larger than $\L^2$. Here the largest eigenvalue is \smash{$\frac{4\pi}{A} N(N+1)$}, and we are let to identify $N(N+1)$ with $A \L^2/(4\pi)$. However, as $N$ is changed, the eigenvalues jump by $\sim 8\pi N/A$, which means that our identification is  unambiguous only up to terms of order $N$ or of order $\sqrt{A}\L$. This kind of subtlety is precisely the reason why the sharp cutoff method does not work in general. Here, let us simply try to identify
\be\label{Nlamrel}
(N+1)^2=\frac{A}{4\pi}(\L^2-M^2)\,\cvp 
\ee
with some arbitrary cutoff-independent constant $M^2$. Note that the form \eqref{Nlamrel} for the scaling is not unique\footnote{
In particular, one could have written $b$ instead of $A M^2$, but in the end this would only shift some constants between the counterterms $\sim A$ and the irrelevant overall normalization constant of the determinant, resp.~partition function.
}, we could have also considered e.g.\ $N(N+1)=\frac{A}{4\pi}(\L^2-M^2)$ which differs by a term $O(N)=O(\sqrt{A}\L)$. Eq.\ \eqref{Nlamrel} will be justified a posteriori by the fact that it yields a good large $\L$ asymptotic expansion. Indeed, with \eqref{Nlamrel} we have 
$$ \ln\Bigl(N(N+1)+\frac{a}{4\pi}\Bigr)=\ln \frac{A\L^2}{4\pi} -\sqrt{\frac{4\pi}{A\L^2}}-\frac{2\pi}{A\L^2}+\frac{a}{A\L^2}-\frac{M^2}{\L^2}+O\Bigl(\frac{1}{\L^3}\Bigr)$$
so that
\be\label{T1-3}
{\cal T}_1\Big(\frac{a}{A}\Big)=
\frac{A}{4\pi}\Big( (\L^2-M^2)\ln\L^2-\L^2\Big) +\Big( \frac{a}{4\pi}-\frac{1}{3}\Big)\ln\L^2\\
+\Big(\frac{a}{4\pi}+\frac{2}{3}\Big)\ln\frac{A}{4\pi}
+ \wt c[a] +O\Big(\frac{1}{\L}\Big)\, ,
\ee
with $\wt c[a]=c[a]+\frac{a}{4\pi}-\frac{1}{2}$.
All terms $\sim\sqrt{A}\L$, which could not be absorbed in counterterms, have cancelled.
Putting things together, we get for the determinant \eqref{zdetN}
\begin{multline}\label{zdetNLam}
\Bigl[\ln\Det'\Bigl( z \bigl(\D_*+a/A\bigr)\Bigr)\Bigr]_N=
\frac{A}{4\pi} \Bigl[(\L^2-M^2)\ln (z\L^2) -\L^2\Bigr]
 +\Big( \frac{a}{4\pi}-\frac{1}{3}\Big)\ln(z \L^2)\\
 +\Big( \frac{a}{4\pi}+\frac{2}{3}\Big)\ln\frac{A}{4\pi z}
+ \wt c[a] +O\Big(\frac{1}{\L}\Big)\, ,
\end{multline}
which is to be compared with \eqref{zdet} for $h=0$, after substituting $t=\alpha/\L^2$. Both expressions have the same structure. While the precise coefficients of the diverging terms are somewhat different, the finite coefficients of $\ln A$ turn out to be exactly the same.

One could then easily continue and compute the quantum gravity partition function in this sharp cutoff scheme on the sphere, i.e.\ $W_1$. Again, it is obvious that there is no multiplicative anomaly in this scheme. The fact that the modes $l=1$ must also be projected out on the sphere and the extra area dependence from the Faddeev-Popov determinant are handled in the same way as in the main text. It is then clear that one does obtain the same $\g_{\rm str}$, i.e.~eq. \eqref{gammastring}.

\end{appendix}


\end{document}